\documentclass{article}

\usepackage{graphicx}
\usepackage{textcomp}

\usepackage{arxiv}

\usepackage[utf8]{inputenc} 
\usepackage[T1]{fontenc}    
\usepackage{hyperref}       
\usepackage{url}            
\usepackage{booktabs}       
\usepackage{amsfonts}       
\usepackage{nicefrac}       
\usepackage{microtype}      
\usepackage{lipsum}
\usepackage{graphicx}
\usepackage{textcomp}
\usepackage{float}
\usepackage{xcolor}
\usepackage{tabularx}
\usepackage{listings}

\title{Software Architecture Metrics: a literature review}

\begin{document}
\author{
  Théo Coulin$^1$, Maxence Detante$^1$, William Mouchère$^1$,Fabio Petrillo$^2$\\
  Département de génie informatique et génie logiciel\\
  Polytechnique Montréal$^1$\\
  Montreal, QC, Canada\\
  \texttt{\{theo.coulin,maxence.detante,william.mouchere\}@polymtl.ca} \\
 \\
  Département de Mathématique et Informatique\\
  Université du Québec à Chicoutimi$^2$\\
  Chicoutimi, QC, Canda\\
  \texttt{fabio@petrillo.com}
}

\maketitle

\begin{abstract}
In Software Engineering, early detection of architectural issues is key. It helps mitigate the risk of poor performance, and lowers the cost of repairing these issues. Metrics give a quick overview of the project which helps designers with the detection of flaws or degradation in their architecture. Even though studies unveiled architectural metrics more than 25 years ago, they have not yet been embraced by the industry nor the open source community. In this study, we aim at conducting a review of existing metrics focused on the software architecture for evaluating quality, early in the design flow and throughout the project's lifetime. We also give guidelines of their usage and study their relevance in different contexts.
\end{abstract}


\section{Introduction}

Building a system in the software engineering field has proven to be a difficult task, especially in large projects. The need to think ahead and to reflect on what the system could look like before starting to implement it is vital. Therefore, the importance of software architecture is critical in any software engineering project. This is the reason why we wanted to explore the quantitative measurements that could be driven on software architectures to check if it has good properties.

The quality of a software's architecture is essential, yet very difficult to apprehend and measure. The quality features of an architecture are not obvious as relations and dependencies can extend very far away. Moreover the complexity and the amount of relations also often make it difficult to read. Evaluating an architecture by hand is a tedious task. This is why we think it is very interesting to have metrics that sum up and quantify the features of an architecture. They help the designer because they hint the flaws to look for.

Having such metrics would make architecture checking much faster and less costly. This would make it possible for designers to run checks from the start of the project \cite{galster_early_2008} and throughout the life of the software. For example, in a scrum context, the architecture could be evaluated at each sprint to make sure it's not drifting to become something impossible to maintain. It would also allow for comparison between architectures to pick the one that fits best to the project's requirements.



The goal of this paper is to provide a systematic review of architecture metrics and to link them with some of the quality features that they represent. Most metrics are not focused on one quality feature, but by combining them, one can end up with a quantitative measure of the quality features.

To reach our goal we will focus on the two following research questions :
\begin{itemize}
    \item \textit{What are the metrics available for a software designer to evaluate the quality of a software architecture?}
    \item \textit{To what extent are these metrics representative of an architecture's quality?}
\end{itemize}


The contributions of this study are twofold. First we will build a reference list of the existing metrics applicable on software architecture. Then we will give guidelines for the usage of these metrics and show what quality features they express and how relevant they are.


The rest of the paper is structured as follows. First, in \autoref{background}, we are going to shortly present the context of software engineering and give definitions for the words we will often use in the paper. Then, we will talk about some related studies in \autoref{related}. In \autoref{design}, we will explain our study design and the methodology we followed. Section \autoref{results-trends} will outline the publication trends of the papers we gathered. Section \ref{results} will present the metrics we identified during our review. For each metric we will give guidelines for usage and check whether the metric can be considered relevant or not. Threats to the validity of our work are considered in \autoref{threats}. Finally we will close the study by summing up the conclusions and thinking about the future work ahead in \autoref{conclusions}.

\section{Background} \label{background}



\subsection{Architecture Qualities}

When designing a software, there are many different ways to come up with the solution to the original problem. In fact, the solution is often found through trial-and-error because in software engineering problems are wicked. The solution to the first problem might unveil something new that will affect your original solution. If it is possible to end up with multiple different design, it is also very likely that only a few of these design are quality design. This is where metrics might first be useful, to quickly determine if a solution is of sufficient quality to be used. Another problem is that the quality of a design can not be quantitatively expressed, as it is a qualitative estimation of a combination of factors whose importance depends on the project's requirements. Here is a list of the quality features of a software architecture that we found during our review :
\begin{itemize}
    \item \textbf{Maintainability} \cite{mo_decoupling_2016,perepletchikov_coupling_2007,perepletchikov_cohesion_2007, dayanandan_empirical_2016} : It is very important that a software is easy to maintain. It must be easy to make changes in a software, either for the addition of a new feature, or for a bug fix. This means that changes must be possible locally, without impacting all of the software. Maintainability also means that it is possible for multiple developers to work on separate parts of the software without impacting each other, which enables parallel work. Maintainability is often synonym with sustainability, as it makes it possible for the software project to be used and updated for many years. For all these reasons, making a software maintainable is key for the future of the project.
    \item \textbf{Extensibility} \cite{dayanandan_empirical_2016} : The ability for a software architecture to handle addition of new functionalities and components. It is very valuable in agile development as features are added throughout the life of the project.
    \item \textbf{Simplicity, Understandability} \cite{alsharif_assessing_2004,zhao_assessing_1998,hofmeister_supporting_2008,mo_decoupling_2016} : Making a software architecture as simple as possible is key to making it most understandable for everyone. Another way to represent this is the complexity of the architecture. Of course the software architecture will always be complex, but the goal is to make it as simple as possible. A simpler architecture is more maintainable, easier to communicate about and possibly more reliable.
    \item \textbf{Re-usability} \cite{qian_decoupling_2006} : In the industry, it is often very valuable to re-use previous projects to make development and design quicker. Making a software architecture re-usable is good to save on future projects.
    \item \textbf{Performance} \cite{williams_performance_1998, distefano_uml_2011} : Having a metric capable of estimating performance early in the design process is very interesting as performance is often an issue in software architecture while it is also often a part of the requirements of the client. This would help with choosing the best architecture for performance before even starting to develop anything, as no implementation trick can compensate for the poor performances of a bad design.
\end{itemize}

\subsection{Relevance}
We will try to estimate the \textbf{relevance} of the metrics we show in this paper. The term relevance has a rather large meaning, so we will try here to explain what criteria we used to assess it. We are aware of the subjective aspect of such a definition, but we tried to highlight what we believe to be key components in the quality of a metric :

\begin{itemize}
    \item \textbf{Relation with architectural qualities} : The metric has to depict efficiently one or more quality features of a software architecture. It seems obvious, but the metric has to provide insight on a given aspect of the architecture.
    \item \textbf{Theoretical validation} : If the metric is complex and uses mathematical concepts, such as statistics for example, its validity has to be demonstrated theoretically.
    \item \textbf{Empirical validation} : The more the metric has been tested in real conditions, the better. A metric, even crafted with the greatest care and proven theoretically, needs to be tested to see if it really provides meaningful information on the architecture.
    \item \textbf{Ease of utilization} : a metric, or a given set of metrics, must to a certain extent be easy to compute. As taking too much time on the architectural phase of a project is prohibitive in business contexts, the time that is necessary to compute metrics has to be reduced to the minimum.
    \item \textbf{Understandability} : A metric should be easy to understand by anyone, as it can be a tool for presenting the advantages and drawbacks of an architecture in a convincing way.
\end{itemize}

\section{Related Work} \label{related}

One of the first papers on the subject \cite{chidamber_metrics_1994}, wrote in 1994 by Chidamber et al., proposed a first set of metrics to apply on a software architecture. Since then, many combinations of their metrics have been studied to produce more relevant indicators. The focus has also been on making possible automatic evaluation of UML designs because a software architecture is not trivial to translate into something understandable by a computer.

We found three Systematic Literature Reviews which deal with software architecture quality metrics.
The first review \cite{koziolek_sustainability_2011} found about 40 metrics distributed over 11 \textit{metrics-based} evaluation methods. But this review includes in its definition of architecture the system implementation. This means that not all the metrics found are relevant in an \textit{architecture-only} context, which makes it a broader review than ours. Moreover, the paper focuses on the sustainability of the software rather than on the quality in general, which makes it look on a different angle than we do. Given that it dates back from 2011 we think that there might have been new metrics or interpretations in the literature.

The second review \cite{stevanetic_software_2015}, which is focused on the \textit{architecture-only} context, found 25 metrics. It builds a mapping of these metrics in the literature, and uses their frequency of appearance to evaluate their maturity. Some metrics we found will overlap with these two other reviews, but we think that we may find different or even new metrics (\cite{stevanetic_software_2015} dates back to 2015). Moreover, our study puts an incentive on the usage of these metrics to improve the quality of an architecture when the others were focused on other aspects.

The third review \cite{staron_portfolio_2017} is a very recent review on software architecture metrics. It focuses on giving software designers tools to evaluate large architectures for industrial applications. They end up building three major categories of metrics : \textit{Architecture Measures, Design Stability and Technical Debt}. Even though they share some common base material with us, their study is different from ours in that they don't show the metrics that they do not endorse as relevant. We, on the counterpart of that, will showcase those as metrics to be tested, as they could lead to an improvement of what is available.


\section{Study Design} \label{design}

We conducted this review applying a methodology and following guidelines from \cite{kitchenham_guidelines_2007} and \cite{kitchenham_systematic_2013}. We put a lot of effort in making it transparent and reproducible. We will explain in the following section the methodology we followed.

\subsection{Research Questions}
Our goal is to answer the following research questions:
\begin{itemize}
    \item \textit{What are the metrics available for a software designer to evaluate the quality of a software architecture?}
    \item \textit{To what extent are these metrics representative of an architecture's quality?}
\end{itemize}

\subsection{Information gathering and interest checking}

As a first approach to find content in the literature we used Google to find some papers about metrics, as using the databases wasn't very effective in the first place. This was our way to get a few papers to verify there was an interest in software architecture metrics and to help us identify the scope of our work. While doing this, we found two SLR, \cite{koziolek_sustainability_2011, stevanetic_software_2015} and \cite{chidamber_towards_1991, chidamber_metrics_1994} which are founding papers for the subject.

\subsection{Tools}
For this review, we used the website parsif.al to keep a trace of everything we did regarding the selection of the papers. These tools helped us having an overview of our paper set and made the selection process easier. We used Zotero to export our bibliography. We also used a lot of shared spreadsheets to divide our work.

\subsection{Search and selection process} \label{methodo:practice}
\begin{figure}[H]
    \centering
    \includegraphics[width=.8\textwidth]{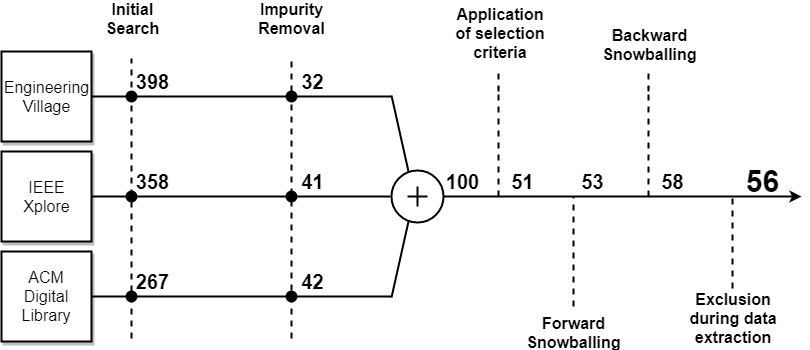}
    \caption{Papers selection methodology and results}
    \label{fig:UC_Create}
\end{figure}

We designed our search and selection process following steps as described in \cite{kitchenham_guidelines_2007,kitchenham_systematic_2013}. It helped us having a broad scope of research but efficiently sorting among the amount of papers available. The steps we followed are depicted in Figure \ref{fig:UC_Create}.

\paragraph{Initial search}
First, we built search queries to dig in known databases, such as IEEE, ACM and Engineering Village.
We selected these databases because they have a very good reputation, and they are very large and complete which makes this review very exhaustive. They also are very well supported by parsif.al and Zotero which made the process of exporting easier.

Our queries were the following :
\begin{itemize}
    \item IEEE : ("software architecture" OR "software architectures") AND metric* AND (evaluation OR quality)
    \item ACM : (+"software architecture" +"software architectures" +metrics +metric evaluation quality)
    \item Engineering Village : (("software architecture" or "SA" ) AND metric* AND (qualit* or eval* or pertinen* or impact or relevanc* ) wn TI) 
\end{itemize}

We applied this query on all metadata for IEEE and ACM, but only on the papers titles for Engineering Village.

\paragraph{Impurity removal}
Third, we performed an impurity removal per database. Given the content of our research, there are many side papers that match the query but are not relevant, ie. papers on code metrics or papers on network architecture. We removed all of these papers to make a coherent set of papers around our subject.

\paragraph{Merging and duplicates removal}
Fourth, we merged the result of each database into a single set and removed the duplicates.

\paragraph{Application of the selection criteria}
Fifth, we performed a three reviewers voting on our dataset to choose the papers that were the most meaningful for our work.
Our selection criteria were the following :
\begin{itemize}
    \item \textit{S1 - Are there any metrics proposed to evaluate the quality of an architecture ?}
    \item \textit{S2 - Is the relevance of the metric assessed ?}
\end{itemize}

\begin{figure}[H]
    \centering
    \includegraphics[width=0.5\linewidth]{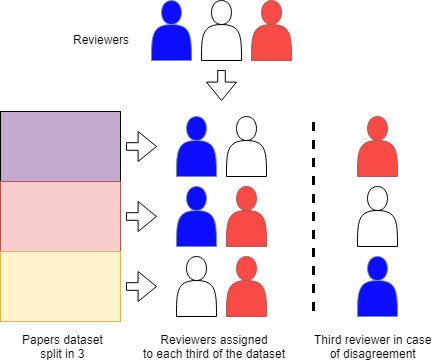}
    \caption{Three reviewers voting for application of the selection criteria}
    \label{fig:reviewers}
\end{figure}

As shown in Picture \ref{fig:reviewers}, we split our dataset in three and assigned two reviewers for each sub-dataset. They went over every paper deciding whether or not it matched our selection criteria. If both reviewers are positive, the paper is in, if both are negative, the paper is out. Finally, if the reviewers did not agree, the third reviewer chose whether to keep or not the paper.

\paragraph{Forward and backward snowballing}
Sixth, we did a forward snowballing using the papers that matches the inclusion criteria, in order to discover more meaningful articles. Seventh, we applied a backward snowballing to make sure that we did not miss any important and founding paper. We performed the snowballing with the following policy:
\begin{itemize}
    \item First we selected only papers which had a title that seemed interesting;
    \item We checked if the paper matched our selection criteria;
    \item We only kept the paper if it was very interesting to us, as we already had a significant database of papers.
\end{itemize}

This means that our snowballing is not exhaustive, but rather focused on adding significant knowledge to our already big set of papers.

\paragraph{Exclusion during data extraction}
While performing our data extraction, there were some papers that we believed were not as relevant as expected. This is why we chose to remove those papers from our final set.

    
\subsection{Data extraction}
The papers that we gathered contain different metrics that we tried to summarize and explain. We also checked whether or not they were relevant.
For every paper, we focused on extracting a few key components that appeared to be be of high importance to us:
\begin{itemize}
    \item Which architectural quality is treated. The extracted data can be a subset of \textit{\{Maintainability, Extensibility, Simplicity and Understandability, Re-usability, Performance\}};
    \item The specific domain or topic associated with the paper. A good example of a domain is Service Oriented Design, but it can also cover a broad topic like change propagation or performance;
    \item The fact that the metric(s) proposed has been theoretically and/or experimentally validated by the researchers.
\end{itemize}

After the data extraction, we synthesized our findings in different tables in order to have an overview of the metrics and then to categorize them by looking at what aspect they were measuring. It allowed us to think about the structure of our paper.

\section{Publication trends} \label{results-trends}

\begin{figure}[H]
    \centering
   \includegraphics[width=.8\textwidth]{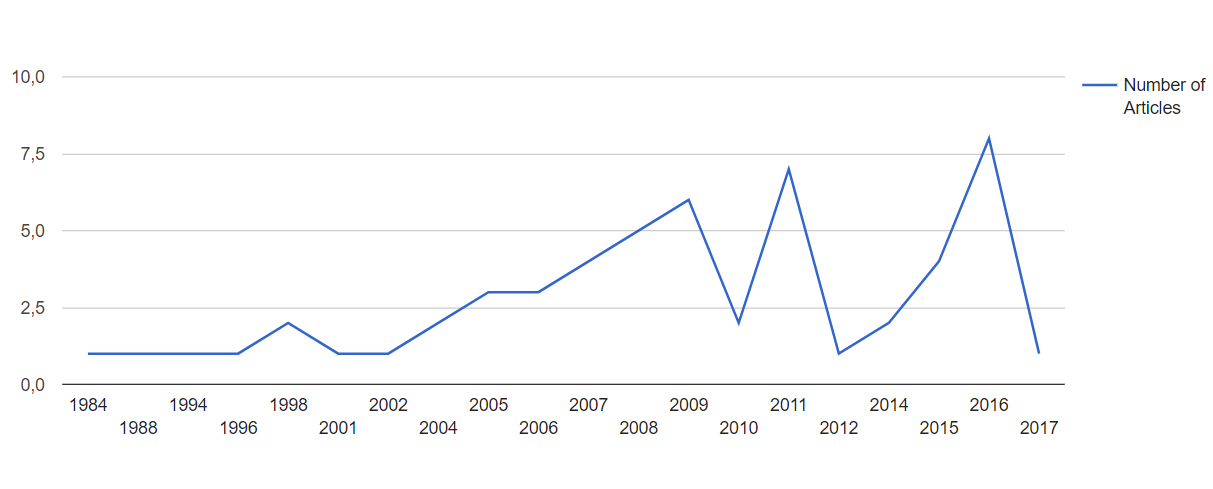}
    \caption{Distribution of publication years in our dataset}
    \label{fig:year-trends}
\end{figure}

We thought that it was interesting to examine a bit the publication trends in our papers dataset, especially the publication years.
Figure \ref{fig:year-trends} shows the distribution of publication years in our dataset.

We can see that software architecture metrics have been a concern for quite some time, as the first paper dates back to 1984. Since then, the scientific interest in this subject has grown a lot to become very trendy in the last couple of years. The industrial interest is surely growing as more and more articles about the subject show a cooperation with the industry \cite{koziolek_morphosis:_2012,mo_decoupling_2016, roubtsov_dn-based_2009}.

\section{Results}
\label{results}

Our research showed that there are groups of metrics which derive from the same underlying measurement. Often papers make little changes in the measurement in order to tweak it and come up with something more meaningful. Sometimes they also make combinations of these measurements. We decided to group the metrics we found by the underlying measurement to which they belong. Then we will present metrics that we found that do not enter in these categories. Finally we will focus on metrics applied to Service Oriented Architecture as they differ a bit from Object Oriented metrics but are very relevant in today's industrial context. As shown by Table \ref{tab:measurements} there are many papers which deal with coupling, cohesion, complexity and size of the software's architecture. We can see a focus on coupling and cohesion, as it is known that it is very important to keep low coupling and high cohesion in an architecture.

\begin{table}
    \centering
    \caption{Underlying architecture measurement}
    \label{tab:measurements}
    \begin{tabularx}{0.47\textwidth}{|l|l|X|}
      \hline
      Metric & \#Papers & Papers \\
      \hline
      Coupling & 28/56 & \cite{ma_evaluating_2009,peldszus_continuous_2016,perepletchikov_coupling_2007,banerjee_quality_2016, shim_design_2008, tvedt_process_2002, alsharif_assessing_2004, galster_early_2008, nicolaescu_evolution_2015, le_relating_2016,hofmeister_supporting_2008,shereshevsky_information_2001,sindhgatta_measuring_2009,venkitachalam_metrics_2015, chahal_metrics_2009, zayaraz_quantitative_2006, choi_dynamic_2007,staron_portfolio_2017,dayanandan_empirical_2016,mo_decoupling_2016, roubtsov_dn-based_2009, serebrenik_dn-based_2009, stevanetic_software_2015, alenezi_software_2016, koziolek_sustainability_2011,sarkar_metrics_2008,basili_validation_1996,perepletchikov_controlled_2011,alahmari_metrics_2011}\\
      & & \\
      Cohesion & 19/56 & \cite{ma_evaluating_2009,peldszus_continuous_2016,perepletchikov_cohesion_2007,shim_design_2008,galster_early_2008,le_relating_2016,shereshevsky_information_2001,sindhgatta_measuring_2009,venkitachalam_metrics_2015,chahal_metrics_2009,zayaraz_quantitative_2006,staron_portfolio_2017,dayanandan_empirical_2016,stevanetic_software_2015,alenezi_software_2016,koziolek_sustainability_2011,perepletchikov_impact_2010,alahmari_metrics_2011,basili_validation_1996}\\
      & & \\
      Complexity & 18/56 & \cite{shim_design_2008, alsharif_assessing_2004,galster_early_2008,banerjee_quality_2016,le_relating_2016,hofmeister_supporting_2008,venkitachalam_metrics_2015,zhao_assessing_1998,beane_quantifying_1984,zayaraz_quantitative_2006,koziolek_morphosis:_2012,staron_portfolio_2017,dayanandan_empirical_2016,stevanetic_software_2015,alenezi_software_2016,koziolek_sustainability_2011,basili_validation_1996,alahmari_metrics_2011}\\
      & & \\
      Size & 14/56 & \cite{peldszus_continuous_2016,shim_design_2008,alsharif_assessing_2004,bouwers_quantifying_2011,ahuja_quantitative_2016,li_towards_2007,chahal_metrics_2009,koziolek_morphosis:_2012,staron_portfolio_2017,dayanandan_empirical_2016,stevanetic_software_2015,alenezi_software_2016,koziolek_sustainability_2011,sarkar_metrics_2008}\\
      \hline
    \end{tabularx}
\end{table}

We also sorted the papers by the quality features they tackle to show what is focused by the literature. This is summed up in Table \ref{tab:features}. We believe that the quality feature the most talked about are very probably features that can now be measured with a metric reliably. This means that these feature are ready to be evaluated automatically on large projects. In the Table \ref{tab:features} we can also see that maintainability is the main focus, probably because it is the most important quality features, as all others depend on it.

\begin{table}
    \centering
    \caption{Quality features tackled by the papers}
    \label{tab:features}
    \begin{tabularx}{0.47\textwidth}{|l|l|X|}
      \hline
      Feature & \#Papers & Papers \\
      \hline
      Maintainability & 31/56 & \cite{perepletchikov_cohesion_2007,perepletchikov_coupling_2007,le_relating_2016,li_towards_2007,christensen_lightweight_2010,sindhgatta_measuring_2009,zimmermann_metrics_2015,venkitachalam_metrics_2015,chahal_metrics_2009,abdelmoez_quantifying_2005,bouwers_quantifying_2014,zayaraz_quantitative_2006,koziolek_morphosis:_2012,perez-palacin_software_2011,choi_dynamic_2007,staron_portfolio_2017,dayanandan_empirical_2016,changjun_architectural_2008,fontana_automatic_2016,shaik_change_2005,mo_decoupling_2016,roubtsov_dn-based_2009,serebrenik_dn-based_2009,stevanetic_software_2015,alenezi_software_2016,koziolek_sustainability_2011,perepletchikov_impact_2010,abdelmoez_error_2004,sarkar_metrics_2008,perepletchikov_controlled_2011,alahmari_metrics_2011}\\
      & & \\
      Performance & 6/56 & \cite{liu_design-level_2005,geetha_framework_2011,distefano_uml_2011,zimmermann_metrics_2015,williams_performance_1998,stevanetic_software_2015}\\
      & & \\
      Extensibility & 6/56 & \cite{sindhgatta_measuring_2009,zayaraz_quantitative_2006,dayanandan_empirical_2016,stevanetic_software_2015,koziolek_sustainability_2011,sarkar_metrics_2008}\\
      & & \\
      Simplicity & 9/56 & \cite{shim_design_2008,li_towards_2007,zimmermann_metrics_2015,staron_portfolio_2017,dayanandan_empirical_2016,fontana_automatic_2016,stevanetic_software_2015,koziolek_sustainability_2011,sarkar_metrics_2008}\\
      & & \\
      Re-usability & 12/56 & \cite{shim_design_2008,qian_decoupling_2006,sindhgatta_measuring_2009,chahal_metrics_2009,abdelmoez_quantifying_2005,bouwers_quantifying_2014,zayaraz_quantitative_2006,staron_portfolio_2017,dayanandan_empirical_2016,changjun_architectural_2008,roubtsov_dn-based_2009,serebrenik_dn-based_2009,stevanetic_software_2015}\\
      \hline
    \end{tabularx}
\end{table}

\subsection{The goal/question/metric paradigm}
First, we want to introduce the work of Basili et al. \cite{basili_tame_1988}. This work is fundamental, as it proposes a framework for creating new metrics. It is based on the goal/question/metric paradigm, which is used in a considerable amount of papers we found.

The aim of this method is to derive metrics that will help assess the quality of a given parameter. In order to be sure that the metric is well-related to the parameter, the framework proposes a three-levels systematic procedure to finally derive software metrics : the goal level, questions level and ultimately metrics level. The contribution of this method is to propose templates for setting goals, and guidelines for deriving quantifiable questions and, finally, metrics.

On the first level, one has to set project goals, which are split into different subcategories : \textit{purpose}, \textit{perspective}, and \textit{environment}.

On the second level, questions must be derived from the goals set. There is also three categories here : \textit{definition of the product}, where questions are related to quality attributes; \textit{quality perspectives of interest}; and \textit{feedback}.

On the last level, deriving metrics from the questions and goals mentioned above, a few guidelines are provided. The difference between subjective and objective metrics is emphasized, where objective metrics are seen as countable characteristics of a product (complexity, for example), and subjective metrics are related to aspects that cannot be characterized objectively. In that case, a ordinal scale may be used to categorize and create a metric.





\subsection{Coupling}

On an architecture level, the metric which is the most used is coupling, and its complementary, cohesion. The first occurrence of coupling as a metric we found was in Chidamber et al. \cite{chidamber_metrics_1994}. It has then been derived into many versions which sharpen the precision of the metric. Generally speaking, what comes from these papers is that it is important to have a high cohesion in modules, and a low coupling throughout the architecture.
Low coupling is very important because it diminishes the risk of ripple effect when making changes in the program. Thus, low coupling is very important to keep the architecture maintainable. Maintainability is the most important quality that is displayed by a low coupling.

In \cite{shereshevsky_information_2001}, Shereshevsky presents a metric for coupling that is applicable on different levels of architecture design. Either at early steps of the design or at code level. They are based on data and information flows. It has not been validated in the paper but seems promising.

In \cite{choi_dynamic_2007}, Choi extends the static coupling defined bu Chidamber to dynamic coupling which is a measurement at object level. The goal is to be even more accurate. It has been compared with static coupling in the paper and seems to give more result, but at the cost of simplicity of use.

Another very interesting approach it that of Mo et al. in \cite{mo_decoupling_2016}. Instead of measuring coupling, they decided to measure the decoupling, that is the modularity of the architecture. A good modularity means easy maintenance and re-usability. Their metric has been validated in the paper and seems very up to date and usable.

\subsection{Complexity}

Another very important metric is complexity. It affects the understandability of the architecture and possibly the performance. It has also been first defined by Chidamber et al. in \cite{chidamber_metrics_1994}.


It can be expressed by the number of classes in the architecture, or the number of links between classes in the architecture. In \cite{zhao_assessing_1998}, the author actually makes a mix of these two values. He calculates the dependence graph of the architecture and uses common graph calculus to assess the complexity of the architecture. This metric hasn't been verified yet.

In \cite{alsharif_assessing_2004}, the author uses Full Function Point, a functional size measure, to calculate the complexity of a software architecture. There is an example of application in the paper, but it has not been widely tested.

\subsection{Change and error Propagation}

Change Propagation evaluates the maintainability of an architecture based on the probability that a change in a class will have an impact on other classes. Shereshevsky et al. \cite{shereshevsky_information_2001} proposed a set of metrics for software architectures in 2001, and introduced change propagation and requirements propagation between components. The difference between the two is that change propagation can help measuring the cost of corrective maintenance, whereas requirements propagation deals with adaption to a change in requirements.

In 2005, Abdelmoez et al. \cite{abdelmoez_quantifying_2005} dedicated a whole paper to change propagation metrics, in component-based architectures. They proposed to build a matrix of change propagation probabilities between components, in order for the architect to be able, at a glance, to assess the difficulty and cost of maintenance operations. Moreover, it can highlight key components that have a high probability to change, or that imply changing other components when they change. Those quick insights allows architects to take counter-measures easily, for example making a component easily adaptable. 

Abdelmoez et al. also worked on metrics related to error propagation in \cite{abdelmoez_error_2004}. It is also based on component-based architecture, introducing metrics to compute the error propagation in a system, using a stochastic approach. They define the error propagation probability between two components, A and B. If there exists a connector between the aforementioned components, we can compute he probability that an error occurring in A is transmitted to B, rather than masked by B.

They also consider the fact that even if a error occurs in A, there is no guarantee that A will transmit a message to B. Therefore, they introduce a transmission probability matrix to assess if an error in A will effectively be transmitted to B, considering not only the probability of the error occurring in A but also the probability of A transmitting a message to B.

Those metrics seem pretty solid on the theoretical part. Moreover, the authors developed a framework to empirically test their metrics by using a fault injection tool. Also, they rigorously compare those results with the analytic ones they got from computing the metrics with only architecture-level information. One can use those metrics to assess if a bug in a given component can impact other component, and therefore gain insight on the future maintainability of a system. 

\subsection{Design Pattern Density}
We found a very promising metric, proposed by Dirk Riehle in \cite{riehle_design_2009}, which measures the percentage of class in the architecture that are part of a design pattern. It helps the designer to evaluate the maturity of an architecture. The more mature an architecture is, the more design pattern are put into it, and the higher the design pattern density.
It is very good when applied on frameworks, which should be very densely filled with design patterns. A framework with high pattern density is more understandable and likely more performing.
This metric seems to be quite hard to use as it does not express on a fixed scale the maturity of the design, but rather on a scale which depends on the problem the software deals with.
It has been tested with success on open source frameworks, but it has not been broadly validated yet and would benefit from being.
We think that this different metric might be a very good way to express maintainability and understandability of the design.

\subsection{Performance Evaluation through architecture}
We also found a few papers related to performance, more particularly Software performance Engineering (SPE), and how to assess the possible performance of a system from its architecture only. It builds on the hypothesis that architecture plays a critical role in the performance of a system.
Liu et al. \cite{liu_design-level_2005} describe their method to assess the performance of a component-based and container-hosted solution. It requires prior modeling of several critical part of the future system, such as the tool that receives all requests and redirects them to the corresponding service, or the database activity. Therefore, it requires additional efforts, but seem to provide an accurate profile of the platform performance through response time prediction.
Distefano et al. \cite{distefano_uml_2011} use UML diagrams and the OMG \textit{Profile for Schedulability, Performance, and Time Specification}. From this point, they construct a performance model, from which they extract several metrics such as utilization and throughput. 

The drawbacks of these methods are the same: the cost of implementing them to compute metrics and solve the models seems to be high and the process time-consuming. Moreover, since performance evaluation is highly correlated with the physical characteristics of the system, the measurements depend a lot on the architect capability to estimate correctly several performance-related factors. Thus, there is a need to have a good knowledge of the performance field to construct the performance model, which will be solved to derive performance metrics. to assess the compliance of the architecture with performance requirements.

Gheeta et al. \cite{geetha_framework_2011} try to implement a lightweight approach to performance evaluation, by only considering \textit{Use Cases} diagrams, and annotating them with performance goals such as a time, or a size, to respect. Then, from technical and environmental factors, and from the complexity of use cases and actors, they estimate a representative workload. From the diagrams made, they generate a performance model, which is then solved to show response times of different components of the system. 


\subsection{Modularization}

In \cite{sarkar_metrics_2008} Sarkar et al. define a set of metrics that is suited to perform an evaluation of large projects' modularity. A project that is very modular is easier to maintain in the long term and can easily be extended. The metrics defined by the author are made to encourage the usage of proper APIs between the modules of the project. Those APIs can be APIs providing service, or extension APIs. The later define a taxonomy of the functionality to provide for the plugins that will extend the module.
Their set includes :
\begin{itemize}
    \item metrics to measure the coupling between modules;
    \item metrics to count the number of inter-module calls that are not made through the defined API;
    \item metrics to detect inheritance between classes of different modules;
    \item metrics to assess that the higher level of abstraction is the one used by classes outside the module (Liskov Substitution Principle);
    \item metrics to assess that interfaces are actually holding a single responsibility.
\end{itemize}
Though dating back from 2008, this metrics set is very modern as it checks some of the SOLID principles. All the metrics have been very thoroughly validated in the paper and are very well described.

\subsection{UML diagrams evaluation}


Li et al. \cite{li_towards_2007} proposed three metrics  to evaluate different aspects of UML diagrams. The first metric proposed is \textit{Information Content} (IC). It is based on a hierarchy and weight of the different elements in UML diagrams. It defines the quantity of information a diagram or the architecture passes. The higher the IC, the higher the amount of information delivered.

The second is more interesting as it is original, and helps assessing the quality of the architecture in terms of understandability. It is called \textit{Visual Effect}. The higher the visual effect, the more complex it is for a human being to comprehend the diagram at a glance.

The third metric is close to being a coupling metric : the \textit{Connectivity degree} measures the number of associations w.r.t. the numbers of entities in the diagram. Different types of associations have different weights.

According to their authors, these metrics can be used to assess the scale, complexity and stability of a given architecture, but these metrics have not been validated experimentally (except a case study).


\subsection{Methodologies for architectural evaluation in agile environment}
We found two papers dealing with controlling software architecture through the whole development process, in the context of agile methodologies. 

Agile methodologies have the particularity to work in iterations. At the start of each iteration, some requirements are chosen by the development team and the client. They become the objectives for the iteration. Those aspects are implemented, and at the end of the sprint, the product is presented to the client. Then, a reflection on the iteration is made to highlights ways to improve, and the process goes on with a new iteration. The advantages of such methodologies are that the product is more resilient in the event of changing requirements. But these methodologies also put emphasis on not having too much documentation.

The problem that can arise is to have a robust software architecture in those conditions. We found two papers that dealt with this paradox of having to keep a robust architecture, while using a methodology that is heavily feature-oriented : Ahuja et al. \cite{ahuja_quantitative_2016} and Christensen et al.\cite{christensen_lightweight_2010}.

The work of Christensen et al. \cite{christensen_lightweight_2010} is very interesting : it starts with the finding that most of the techniques used to assess the quality of software architectures are heavyweight and costly to perform, thus not useful in an agile context. They developed the \textit{architectural Software Quality Assurance} technique (aSQA) to provide a lightweight technique, whose goal is to assess the quality of software architecture as well as prioritizing things to work on. It also enables to balance quality attributes of the architecture. Also, this technique needs to have a component based, or service-based architecture to be easy to use.

The most important part of the technique lies in the beginning of its implementation. At the start of each project, every stakeholder has to agree on the qualities the product should have. This allows to derive architecture qualities. The next step is to define metrics to quantify the quality attributes of the architecture. 

The next step is to define a mapping of quality measurements to aSQA levels. Since the aSQA technique puts emphasis on balance and prioritization between qualities, it is necessary to use the same scale for all metrics. Therefore, they decided to use ordinal values between 1 and 5 to allow comparison between quality attributes. 1 defines an unacceptable level of the quality attribute, 3 an acceptable one, and 5 an excellent level of quality.

From those metrics, a target level and the current level of the quality attributes are assessed. It enables to compute the health of the quality attribute, and along with its importance (defined by the stakeholders), we can derive the focus to put on this particular quality.

The advantage of this technique is also that it has been tested as the reference technique in a danish company for multiple projects, and also in two other companies. Therefore, its validity is acceptable.

We can conclude that using metrics to assess the quality of software architecture can be interesting even when using processes that do not normally put an emphasis on architecture, as it allows to balance qualities of the product, and also to prioritize and focus attributes for the next iteration.

\subsection{Service Oriented Architecture}
A designer working on a Service Oriented Architecture (SOA) could think that using Object Oriented (OO) metrics is a good way to evaluate his design. However, it as been proven that OO metrics are particularly irrelevant for assessing the quality of a SO design, as concepts of classes, methods or objects are totally nonexistent in SOA\cite{shim_design_2008}. New metrics have to be derived from the OO ones to adapt their measures to the SOA concepts, like procedures for instance. This is the reason why we decided to dedicate a whole sub-section to the SOA oriented metrics that we reviewed during our work.

Considering Web services in particular, Qian et al. proposed in 2006 a (de)coupling metric\cite{qian_decoupling_2006} for evaluation in service composition of service oriented components. They proposed four metrics giving assumptions of how their evolution impact different quality attributes such as maintainability and understandability. However, no validation of the metrics are provided.

Perepletchikov et al. introduced cohesion metrics especially for SOA in \cite{perepletchikov_cohesion_2007}. They define some characteristics and identify types of cohesion with a ranking to determine which one is stronger than the other. They present a set of five metrics, evaluating the different types of coupling with a sixth one for assessing the total cohesiveness of the software. Their work is particularly interesting as the metrics are calculated at design time and are thus technology independent. On the other hand, there is no empirical validation of their set in this paper, but they plan it as future work.

In a second work the same year, the same authors created a second set of metrics for evaluating the coupling of a SO design\cite{perepletchikov_coupling_2007}. They made eight assumptions about different types of coupling to help them define a set of metrics covering these assumptions. They introduce eight primary metrics and six aggregation metrics based on computations on the primary ones. Note that we decided to consider only their purely static metrics as the other are dynamic metrics, evaluated during run-time, which is out of our scope. Despite not having a real empirical validation, they used the \textit{property-based software engineering measurement framework} referenced in their paper to validate the presented metrics. As their previous work, those are design-time metrics, allowing for maintainability prediction from high level of abstraction in the architectural model. The importance of their metrics is not yet established and they leave it for future work.


Finally, the year after, Perepletchikov et al. conducted a similar study\cite{perepletchikov_controlled_2011} aimed at the validation of the SOA coupling metrics that they introduced in their previous work\cite{perepletchikov_coupling_2007}, through the study of the impact of coupling on the maintainability of an architecture. The paper is organized like the one we previously presented (\cite{perepletchikov_impact_2010}. The experimental protocol used for the validation is very similar to the first one, and we suppose that the participants of this study are the same. The results show that there is a need of more empirical validation, with more case studies and participants, as the weights of the metrics are not well established. In addition, the validation has shown that some types of coupling (e.g. intraservice coupling) have a lesser impact on maintainability than other ones (e.g. indirect and direct, incoming and outgoing, extraservice coupling). Sadly, they do not propose the same guidelines as in their previous work, but this study is a very interesting complement of the early effort they put in coupling metrics \cite{perepletchikov_coupling_2007}.

Hofmeister and Wirtz introduced a complexity\cite{hofmeister_supporting_2008} metric from the coupling point of view. They define six base measures, then three metrics for measuring the coupling, based on the previous ones and finally three metrics for measuring the complexity. For each of them, they give explanation on how they computed it and discuss the exact signification of what is measured. Before testing their model on a case study, they state that their metrics are relevant only if they give a bad rating of a software. In this case, a redesign could be considered, but if the metrics state that there is no need for a redesign, it does not mean that the model is optimal. In addition they do not consider the categorization of a design between "easy to modify" and "hard to modify" feasible. That's why they only introduce these metrics and specify that there is a need of studying their behavior on more "enterprise-scale case studies".

In 2008, Shim et al. came with a very robust paper\cite{shim_design_2008} dealing with SOA metrics in general. First, all of their metrics are evaluated at design time, before any implementation work begin. Second, they extend QMOOD, a hierarchical quality assessment model for object-oriented design quality evaluation, to adapt it to SOA evaluation. The resulting model is a four-level meta-model with 1) High-Level Design Quality Attributes, which were also defined by them, 2) SOA Design Properties, 3) SOA Design Metrics and 4) SOA Design Components. In addition they clearly explain the mappings between each level in the model. Moreover, they define twenty-one basic metrics and seven derived metrics, computed on the previous ones and they establish relationships between these derived metrics and the SOA Design Properties. Finally, they link the properties with the High Quality Attributes by assessing the impact of each property on the quality attributes. At the end of their study, they validate their model by an empirical study on two versions of a project, in order to valuate the quality of the changes. This work is a real milestone in the SOA architecture metrics field. We think that any SO architect should have a look at their model if he wants to evaluate his design and truly understand the results of the analysis.

Ma et al. proposed an interesting study \cite{ma_evaluating_2009} on evaluating the identification of SOA services. As this identification is one of the most important tasks in defining an SOA, is could be a good thing to have a way to quantitatively measure the quality of an identified services portfolio, in order to further be able to compare multiple set of services and eventually choose the best one. Again, as this study focus on service identification, it focuses on the design level of the development process. As other SOA studies, they first define a set of four specific SO metrics to fulfill the specific needs of such a design. Each metric evaluate respectively : Service granularity, Service coupling, Service cohesion and Business entity convergence. The proposed model is conducted in three steps: 1) Modeling, identify services in the business process and model the structure of the identified portfolio; 2) Measuring, use the model to measure quality features of services in the portfolio with the help of corresponding design metrics; 3) Evaluation : overall evaluation of the services set by normalizing the metrics and adding some weights, which any designer can customize to adapt the model to his needs. Finally, the model is validated by a case study on five different portfolios of identified services for a same problem. The evaluation revealed that the model seems quite accurate as the most balanced portfolio was selected as the best in this case.

Sindhgatta et al. defined a rather large set of metrics\cite{sindhgatta_measuring_2009} for evaluating many different quality aspect of a SOA, namely service cohesion, coupling, re-usability, composability and granularity. Their study is particularly interesting because they put a real effort in evaluating and measuring the validity of their metrics, for each category, through two industry projects. In addition, these metrics are calculated and based on the design level only, allowing for early detection of design flaws. This paper is a very good way for an architecture designer to get into SOA evaluation as the authors present many different quality aspect, study the correlations and links between their metrics and finally confront their theory to two real case study, giving range of values for each metric which helps for comparison.

A second paper that deals with service granularity as been proposed by Alahmari et al.\cite{alahmari_metrics_2011} in 2011. In this study, they propose a framework based on eight metrics. In addition of presenting each metric, they also explain what they exactly evaluate and proceed on a theoretical validation using Briand et al. framework (widely used in the metrics sector). They end their paper with a test of their model on a case study. This test show that "the   granularity   of   service   operations always affects complexity, cohesion and coupling" but the exact extent to what it affects those aspects is not yet established. We see two main issues with this paper : first, there is not real empirical validation of this set of metrics, and second, the metrics are based on the code syntax, thus it is difficult to really use them to make decision during the design process. However, they can be useful in a refactoring process for flaws identification.


\section{Threats to Validity} \label{threats}

\textbf{External validity.}

Even though we selected the papers for this literature review with the greatest care, there might be articles that are biased, or for which the experimentation process is not rigorous, as the metrics are tested on a few projects only. This might lead to some metrics that are misleading or not that relevant. 

\textbf{Internal validity.}
There is a great threat on this particular aspect. Since we conducted this review in less than three months, and worked on it as a project while attending multiple courses, there is a threat on all levels of our analysis. From the methodology to the results, the lack of proper time might induce some flaws in the design or in the thought process. For instance, during the forward and backward snowballing, we selected the papers mainly with their title, which are sometimes misleading. Consequently, we may have missed important publications, as we dug down only if the title seemed relevant. Nevertheless, we tried to follow the best practices in SLR, as described in \cite{kitchenham_guidelines_2007,kitchenham_systematic_2013}.

Also, by focusing on the architecture quality, we are missing a lot of quality properties of a software product. We are also letting many of these properties to the implementation phase and are exposed to deviation from the original architectural design. However, we deliberately reduced our scope to the architecture level as we wanted to focus on the design flaws rather than the implementation flaws. Less than a real threat, it is something that one have to have in mind when considering evaluating his software following the guidelines and metrics we reviewed in this paper.

\section{Conclusions and Future work} \label{conclusions}
We conducted a Systematic Literature Review to find which metrics were available to assess the quality of an architecture, early in the design process and throughout the software's lifetime.
It has been known for a while that architecture is a key component in software development. Therefore, we listed a wide range of metrics, sorting them by the quality aspect of architecture they tackle.
We aimed at giving architects an overview of the tools available, in order for them to have a better understanding and another viewpoint when choosing between different architectures during the early stages of a project.
In addition, the cost-effectiveness of metrics allows designers to quickly and quantitatively evaluate their architecture, leaving an open window to comparison. This is a real improvement considering the slow and qualitative evaluation that can be given by expert only, which can not really be integrated in a fast development process like scrum for instance.


Future work may focus on assessing the robustness of the metrics mentioned in this paper. For example, trying to get more experimental data to confront the validity of the metrics might be interesting, as they are not always tested on real projects. This could help the adoption of those metrics by the industry, as they seem to be not that relevant in their point of view\cite{nicolaescu_evolution_2015} for now.

Most evaluation tools, like SonarQube\footnote{\url{http://docs.sonarqube.org/display/SONAR/Metric+definitions}} do not implement any metrics that are design based. These tools put the incentive on the codebase and we believe that it would be beneficial to add metrics that are at design-level rather than at code-level.

Finally, an interesting subject to put effort in is to use some of those metrics and try to employ them together in order to detect other flaws, like security weakness or components identification\cite{nicolaescu_evolution_2015}

\bibliographystyle{IEEEtran}
\bibliography{biblio}

\end{document}